# Infrared nanoscopy of Dirac plasmons at the graphene-SiO$_2$ interface

Zhe Fei[1*], Gregory O. Andreev[1*], Wenzhong Bao[2], Lingfeng M. Zhang[1,3], Alexander S. McLeod[1], Chen Wang[4], Magaret K. Stewart[1], Zeng Zhao[2], Gerardo Dominguez[5], Mark Thiemens[4], Michael M. Fogler[1], Michael J. Tauber[4], Antonio H. Castro-Neto[6], Chun Ning Lau[2], Fritz Keilmann[7], Dimitri N. Basov[1]

[1]Department of Physics, University of California, San Diego, La Jolla, California 92093, USA

[2]Department of Physics and Astronomy, University of California, Riverside, California 92521, USA

[3]Department of Physics, Boston University, Boston, Massachusetts 02215, USA

[4]Department of Chemistry and Biochemistry, University of California San Diego, La Jolla, California 92093, USA

[5]Department of Physics, California State University, San Marcos, California 92096, USA

[6]Graphene Research Centre and Department of Physics, National University of Singapore, 117542, Singapore

[7]Max Planck Institute of Quantum Optics and Center for NanoScience, 85714 Garching, Germany

**ABSTRACT:** We report on infrared (IR) nanoscopy of 2D plasmon excitations of Dirac fermions in graphene. This is achieved by confining mid-IR radiation at the apex of a nanoscale tip: an approach yielding two orders of magnitude increase in the value of in-plane component of incident wavevector $q$ compared to free space propagation. At these high wavevectors, the Dirac plasmon is found to dramatically enhance the near-field interaction with mid-IR surface phonons of SiO$_2$ substrate. Our data augmented by detailed modeling establish graphene as a new medium supporting plasmonic effects that can be controlled by gate voltage.

Surface plasmons are fundamental collective modes of electrons that enable functionalities at the intersection of nanophotonics and electronics[1-5]. Dirac plasmons of graphene, which are the density waves of Dirac fermions, are predicted to enable both low loss and efficient wave localization up to mid-IR frequencies[6-10]. Theoretical studies show that the combination of tunability and low loss is highly appealing for implementation of nanophotonics, optoelectronics, and transformation optics based on Dirac plasmons[9-12]. Electron-energy-loss spectroscopy studies of epitaxial graphene on SiC substrate verified plasmonic effects[13-15]. So far, optical phenomena associated with surface plasmons of the massless quasi-particles in graphene have remained unexplored. This is in part due to the difficulty of carrying out IR experiments at wavevectors matching those of plasmons, which are beyond the reach of conventional transmission or reflection measurements[16]. To overcome this limitation, we employed scattering-type scanning near-field optical microscope (s-SNOM). Previously, this technique was widely applied to studying surface phonons and phonon polaritons[17,18]. In this letter, we identified spectroscopic signatures attributable to the Dirac plasmon and its interaction with the surface phonon of the $SiO_2$ substrate. Our work affirms the under-exploited capability of tip-based optical nanoscopy to probe collective charge modes far away from $q \sim 0$ of conventional optical spectroscopy.

In our experiments, we utilized a commercial s-SNOM (NeaSNOM, neaspec.com) coupled to several interchangeable lasers: two quantum cascade lasers (daylightsolutions.com) and two $CO_2$ lasers (accesslaserco.com). These lasers allow coverage of the mid-IR region from 883 to 1270 cm$^{-1}$ (Figs. 1,2). This region accommodates characteristic features of the electromagnetic response of monolayer graphene[19-21] along with vibrational modes of $SiO_2$. The IR nanoscope is built on the basis of an Atomic Force Microscope (AFM) operating in tapping mode. We acquired near-field images with tapping frequency $\Omega \sim 270$ kHz and tapping amplitude $\Delta z = 40$ nm at ambient conditions. The back-scattered signal is demodulated at the 2$^{nd}$, 3$^{rd}$ and 4$^{th}$ harmonics of the tapping frequency yielding background-free images[22]. The scattering amplitude $s$ and phase $\phi$ at all harmonics are obtained simultaneously with AFM topography by pseudo-heterodyne interferometric detection[23].

Figure 1a displays a schematics of the nanoscopy experiment. The beam of an IR laser is focused on the metalized tip of an AFM cantilever. The strong near-field confinement of mid-IR radiation at the tip apex has two principal effects. First, the collection of back-scattered light from a confined volume characterized by the tip radius $a$ enables IR imaging/spectroscopy at sub-diffractional resolution[22]. Second, the light-matter interaction induced at the vicinity of the tip peaks for in-plane momenta $q \approx 1/a$ far beyond the light line given by $q = \omega/c$. It is this combination of high spatial resolution and high-$q$ coupling that enables us to investigate the spectroscopic signatures of Dirac plasmons by means of IR nanoscopy.

Graphene samples were fabricated by mechanical cleavage of graphite and then transferred to the surface of a 300 nm thick $SiO_2$ on a Si wafer. Commonly, graphene on $SiO_2$ is characterized by a rather high carrier density due to unintentional doping[24,25]. Raman spectroscopy[26] was used to select monolayer samples with nearly identical hole doping $n = (2.9 \pm 1.0) \times 10^{12}$ cm$^{-2}$ corresponding to a chemical potential of $|\mu| = (1600 \pm 300)$ cm$^{-1}$ that we determined as $|\mu| = \hbar$ ($v_F \approx 1 \times 10^6 m/s$ is the Fermi velocity of graphene). The uncertainty in the estimate of the graphene chemical potential is due to ambiguities in Raman measurements of the carrier density as well as data for the Fermi velocity of graphene. For the purpose of absolute spectroscopic measurements, we etched off $SiO_2$ in several regions of the wafer to access the Si surface. The near-field response of Si is frequency-independent in the mid-IR region. Therefore, Si can serve as a convenient reference for a quantitative analysis of the nanoscale electrodynamics of graphene on $SiO_2$. For gating experiments, we fabricated electric contacts to the graphene surface. By varying back-gate voltage $V_g$, we are able to tune the carrier density of graphene $n = C_g \times |V_g - V_{CN}|/e$, where $C_g = 115$ aF μm$^{-2}$, and $V_{CN}$ is the gate voltage corresponding to the charge neutral graphene.

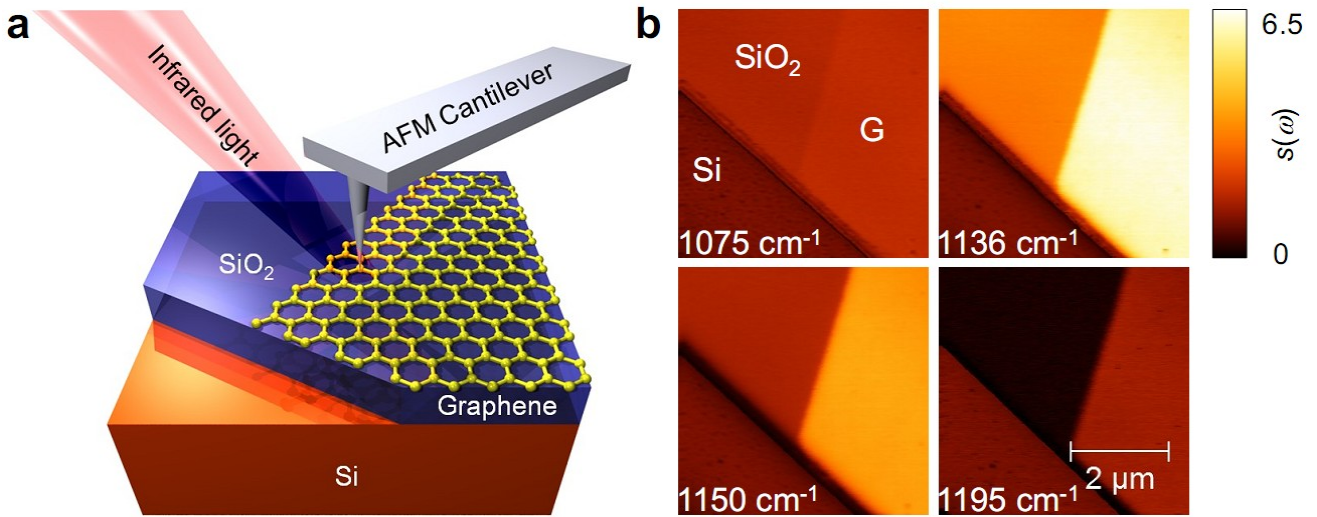

Fig. 1. (a) Schematics of the near-field nanoscopy experiment used to study monolayer graphene on top of $SiO_2$/Si substrate. In the bottom left corner of the structure, $SiO_2$ has been etched away to enable tip contact with Si wafer. (b) Infrared near-field images displayed at four representative frequencies. The strong IR contrast between Si, $SiO_2$ and graphene (G) is clearly seen to vary systematically with the probing frequency. The plotted quantity $s(\omega)$ is the normalized backscattering amplitude defined in the text.

In Fig. 1b we show representative IR nanoscopy images, in which we plot the backscattering amplitude normalized to Si: $s(\omega) = s_3(\omega)/s_3^{Si}(\omega)$. Here, the backscattering amplitude $s_3(\omega)$ is demodulated at the 3$^{rd}$ harmonic of the tip tapping frequency. The simultaneously recorded AFM topography is displayed in Fig. S1 in the Supporting Information. These images reveal nearly uniform signals in either SiO$_2$ or graphene regions with characteristics varying systematically with IR frequency. In Figs. 2a,c we present these results in the form of both amplitude $s(\omega)$ and phase $\phi(\omega) = \phi_3(\omega) - \phi_3^{Si}(\omega)$ spectra. Each data point in Figs. 2a,c was obtained by averaging over corresponding areas in images similar to those displayed in Fig. 1b.

We first consider the amplitude spectra of SiO$_2$ which reveal a near-field resonance centered at $\omega = 1128$ cm$^{-1}$ due to the surface phonon of SiO$_2$ in accord with the earlier data[27,28]. The dominant feature of the $s(\omega)$ spectrum for graphene on SiO$_2$ is similar to that for SiO$_2$ alone. However, the most surprising finding is that graphene strongly enhances the amplitude $s(\omega)$ in the 1110 - 1250 cm$^{-1}$ spectral region and also blue-shifts the peak frequency by about 10 cm$^{-1}$. We hypothesize that both effects are related to the high density of mobile carriers present in our graphene samples. In order to verify this hypothesis, we monitored the evolution of the resonance with gating voltage $V_g$ that enables controlled variation of the carrier density in graphene. In the inset of Fig. 2a, we show the results of gating experiments performed at $\omega = 1150$ cm$^{-1}$ where graphene-induced enhancement of the scattering amplitude is most prominent (160% compared to SiO$_2$). At negative $V_g$ corresponding to the increase of the hole density, the scattering amplitude is further enhanced (up to 300% at $V_g$ = -50 V). On the contrary, positive $V_g$, which reduces the hole density, significantly suppresses the contrast between graphene and the oxide. The contrast is minimized at $V_g = (40 \pm 5)$ V, which we assign to charge neutrality voltage $V_{CN}$. This estimate of $V_{CN}$ is in accord with the Raman probe of the carrier density in ungated graphene layers (see Supporting Information). In addition, graphene induces a steep increase of the near-field phase below 970 cm$^{-1}$ (Fig. 2c). We will show that the latter effect stems from direct interaction of ultra-localized IR light with the Dirac plasmon whereas $SiO_2$ resonance modifications originate from plasmon-phonon coupling at the graphene-$SiO_2$ interface.

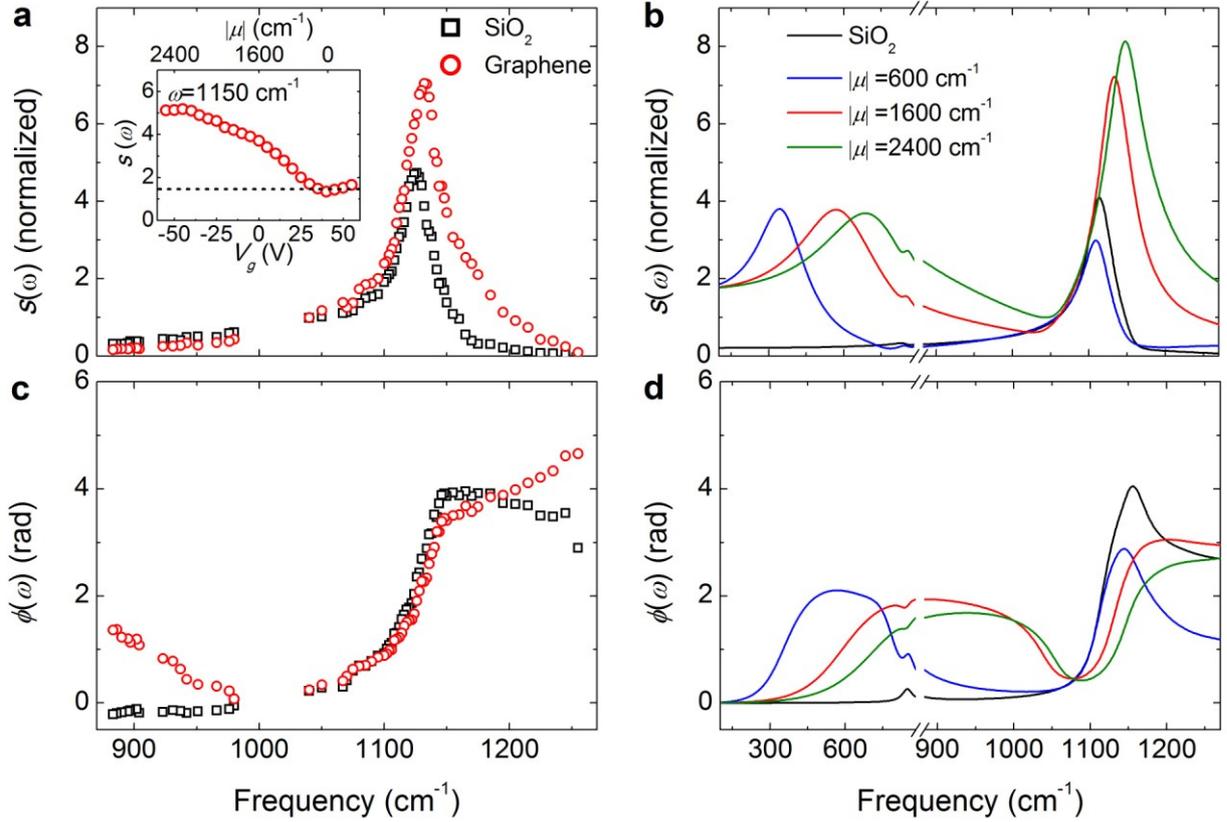

Figure 2. Spectra of the near-field amplitude $s(\omega)$ and phase $\phi(\omega)$. Panels (a),(c): experimental data extracted from images as in Fig. 1b for $SiO_2$ (black squares) and graphene on $SiO_2$ (red circles). The inset of Fig. 1a shows gating measurement results for the graphene near-field amplitude at $\omega = 1150$ cm$^{-1}$. The dotted line marks the value of gate-independent $SiO_2$ amplitude also probed at 1150 cm$^{-1}$. Top axis of the inset marks the calculated chemical potential of graphene. Panels (b),(d): dipole model spectra for $SiO_2$ (black) and graphene on $SiO_2$ (colors) for three different choices of the chemical potential $|\mu|$ = 600, 1600, and 2400 cm$^{-1}$; the predicted low-frequency resonance, the onset of which is seen in (c), reveals direct near-field coupling to the Dirac plasmon of graphene

The physics of the near-field interaction is that the tip, polarized by incident IR light, gives rise to evanescent fields with a wide range of in-plane momenta $q$. When the tip approaches a polar and/or conducting surface, the evanescent fields are altered which in turn affects the tip polarization.

To quantify this interaction we introduce the reflection coefficient, $r_P(q,\omega)$, defined as the ratio of the amplitude of the *P*-polarized reflected field $E_r$ to that of the *P*-polarized incident field $E_i$. This frequency- and momentum-dependent response function completely describes the electrodynamics of the graphene-SiO$_2$ interface, not only in the near field, but also in the far field (see Supporting Information):

$$r_P(q,\omega) = \frac{\varepsilon_1 k_0 - \varepsilon_0 k_1 + (4\pi k_0 k_1 \sigma/\omega)}{\varepsilon_1 k_0 + \varepsilon_0 k_1 + (4\pi k_0 k_1 \sigma/\omega)} \quad (1)$$

In Eq. 1, $\varepsilon_0$ is the dielectric constant of vacuum, $\varepsilon_1$ is the complex dielectric function of SiO$_2$, $k_j = \sqrt{\varepsilon_j(\omega/c)^2 - q^2}$ are the out-of-plane components of momenta, and $\sigma = \sigma(q,\omega)$ is the in-plane optical conductivity of graphene that was obtained from the Random Phase Approximation (RPA) method (see Supporting Information). From Eq. 1, one can see that the in-plane properties of graphene are responsible for its response in sSNOM experiments. This is due to the radial component of the tip's scattered field, which drives charges within graphene. Likewise, these charges impact the tip polarization in response. Note that $r_P(q,\omega)$ diverges at $q$ and $\omega$ values given by the dispersion of the two surface modes at the graphene/SiO$_2$ interface: the SiO$_2$ surface phonon at 1128 cm$^{-1}$ and Dirac plasmon of graphene. A formal connection between $r_P(q,\omega)$ and the direct experimental observable of IR nanoscopy, $se^{i\phi}$, is worked out in the Supporting Information (Eq. S3) by modeling the apex of the tip as a point dipole. Here we only briefly comment on the essential aspects of the modeling procedure. An important parameter of our point-dipole model is the AFM tip radius $a$, which we have set at $a = 30$ nm according to the specifications of our cantilevers. The tip radius determines the effective dipole polarizability $a^3$. Another significant parameter $b$ is the distance between the point dipole and the apex of the tip. Finally, we stress the central result of the dipole-model analysis: the near-field coupling integral $G$ has the weight function of the form $q^2 \exp(-2qz_d)$, where $z_d = b + \Delta z(1-\cos\Omega t)$ is the distance between the tip dipole and the sample surface (Eq. S2 in Supporting Information). The magnitude of $z_d$ is varying with time due to tip tapping. The plot of the time-averaged weight function $\langle q^2 \exp(-2qz_d) \rangle_t$ reveals a bell-shaped momentum dependence that peaks around $q = 1/a$ (Fig. 3a). Thus the dominant in-plane momenta contributing to near-field coupling are distributed around $q = 1/a$ (dashed line in Figs. 3a-d). For that reason, the $s(\omega)$ spectra show resonances if and only if the dispersion curve of a mode intersects the dashed line that marks the dominant near-field momentum. For a typical value of our tip radius, $a = 30$ nm, the probing in-plane momentum exceeds that of the incident light at $\omega \sim 1000$ cm$^{-1}$ by about two orders of magnitude. These virtues of tip-enhanced near-field coupling enable the exploration of both the Dirac plasmon of graphene and plasmon-phonon coupling, which are fundamentally finite-momenta effects.

The dipole model of the near-field interaction[17,27-30], which we have adapted to the graphene-SiO$_2$ interface and augmented with the explicit account of the high-momentum coupling, reproduces all aspects of the data (Figs. 2b,d). We first consider the near-field spectra of SiO$_2$. Comparing the results of dipole-model calculations with measurements, we find near quantitative agreement. Despite overall agreement between the data and modeling, one witnesses minute discrepancies that may stem from two main factors. First, we used bulk optical constants of SiO$_2$ extracted from far-field ellipsometry measurements of our wafers in modeling the surface response (see Supporting Information). Second, the point-dipole model neglects the actual geometry of the tip that may introduce finite dipole or even higher multi-poles to the near-field interaction[28,31,32].

We now proceed to describe the dipole-model results for graphene on SiO$_2$. In Figs. 2b,d we plot spectra of both amplitude and phase, displaying the evolution of the near-field response with variations in the chemical potential $\mu$. For the specific choice of $|\mu| = 1600$ cm$^{-1}$ (consistent with our Raman measurements and also gating experiments), we find that the model spectra reproduce the key characteristics of the data: enhancement of the resonance and its blue shift. The net result is that the Dirac plasmon of graphene radically modifies the SiO$_2$ surface phonon response, which is the experimental manifestation of the plasmon-phonon interaction at the graphene/SiO$_2$ interface.

In order to map the dispersion of the plasmon, we evaluated the divergence of $r_P(q,\omega)$ using Eq. 1 (Figs. 3c-d). The dispersion of the Dirac plasmon approximately follows the square-root $q$-dependence $\omega_P(q) \propto \sqrt{k_F q}$ for $q$ values smaller than the Fermi wavevector $k_F$[33]. Moreover, the plasmon frequencies are also governed by the chemical potential or carrier density $n$ in the graphene layer since $k_F = \sqrt{\pi n}$. The value of the chemical potential also determines the onset of interband transitions and cut-offs for intraband excitations[20] (white dotted lines in Figs. 3b-d). Within the RPA approximation and considering constant scattering rate of quasiparticles in graphene (due to phonons or impurities), the chemical potential alone defines the optical conductivity of graphene in the mid-IR region (see Supporting Information)[19,20,34]. In weakly doped graphene, the Dirac plasmon and the surface phonon of $SiO_2$ are well separated from each other (Fig. 3b). Both modes can be excited in the near-field experiment since their dispersion curves fall within the momentum range of the probe. We note here that surface phonons are extremely localized in real space: a product of their nearly flat dispersion. In contrast, the Dirac plasmon of graphene is a propagating mode, and the real-space aspects of this plasmon will be a subject of future imaging experiments. At moderate levels of carrier density, the plasmon approaches the surface phonon of $SiO_2$ leading to the familiar effects of mode repulsion and hybridization (Fig. 3c). Increasing the carrier density further leads to drastic changes in the dispersion of both the plasmon and the surface phonon (Fig. 3d). In panels Figs. 3b-d one can also notice a structure near 850 cm-1 originating from the plasmon coupling to a weaker low-frequency phonon mode of $SiO_2$.

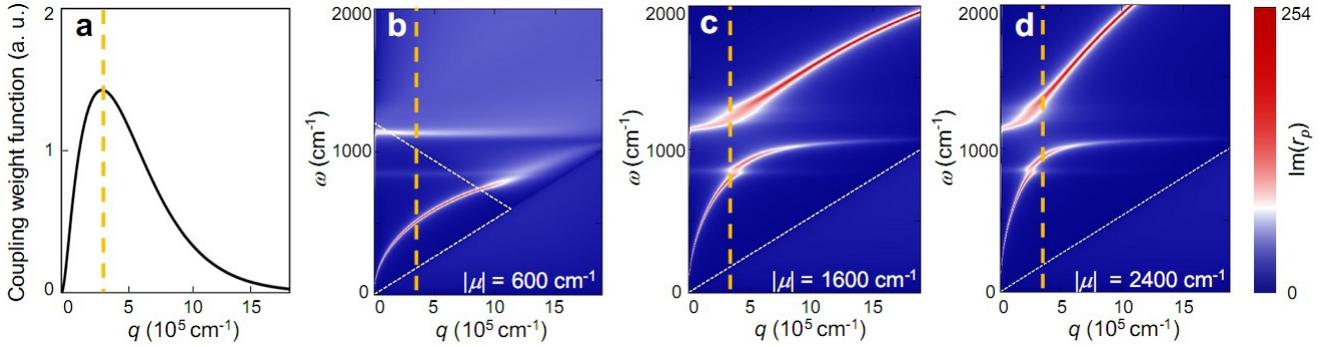

Figure 3. (a) The momentum dependence of time-averaged near-field coupling weight function $\left\langle q^2 \exp(-2qz_d) \right\rangle_t$ which peaks at $q = 3.4 \times 10^5$ cm$^{-1}$ for our tip radius $a = 30$ nm. (b)-(d) Imaginary part of the reflection coefficient $r_P(q, \omega)$ calculated using Eq. 1 with chemical potentials $|\mu| = 600$, 1600, and 2400 cm$^{-1}$, respectively and displayed in false color scale. Vertical yellow dashed lines in (a)-(d) mark the dominant $q$ for maximum near-field coupling. White dotted lines in (b)-(d) mark the boundaries of single-particle intra- and inter-band excitation continua of graphene; these two boundaries meet at ($q = k_F$, $\omega = |\mu|/\hbar$).

The dipole model predicts that the plasmon-phonon interaction and hybridization at the graphene-SiO$_2$ interface (Figs. 3b-d) can be readily observed by near-field nanoscopy. We focus again on the chemical potential $|\mu|$ = 1600 cm$^{-1}$. The dipole model calculations show that the anti-crossing of the Dirac plasmon and the phonon not only causes the blue shift of the peak in $s(\omega)$ spectra but also increases the strength of the resonance (Fig. 2b). Both effects were observed by our experiment. Furthermore, the model predicts the systematic variation of the scattering amplitude with the chemical potential in the 1100 - 1250 cm$^{-1}$ range, which was observed by our gating experiments (inset of Fig. 2a). Because graphene on SiO$_2$ is unintentionally doped, the enhancement of $s(\omega)$ is expected to show an non-monotonic variation with the gate voltage, and have the minimum near charge neutrality point. This is also in accord with the data presented in the inset of Fig. 2a. In combination, near-field spectra in Fig. 2 and gating data at a selected probing frequency attest to the hybrid character of the resonance, involving coupled plasmon-phonon oscillations which dominate the mid-IR response of the graphene-SiO$_2$ interface. One can also anticipate a hardening of the phonon resonance of SiO$_2$ due to screening associated with mobile charge in the graphene layer, a complimentary viewpoint on the effects reported in Fig. 2a (Supporting Information).

Yet another salient feature of the modeled spectra is a strong near-field resonance close to the low-energy cut-off of our data (Figs. 2b,d), which originates from the direct near-field coupling to the Dirac plasmon of graphene. This low-frequency resonance is clearly broadened compared to the hybrid plasmon-phonon mode discussed above. A detailed discussion of the linewidth of both modes is provided in the Supporting Information. Resonance structure due to the Dirac plasmon is clearly visible both is in the $s(\omega)$ and $\phi(\omega)$ spectra (Figs. 2b,d); these features systematically shift to higher frequencies with increasing doping level of graphene. For $|\mu|$ = 1600 cm$^{-1}$, the amplitude resonance of the Dirac plasmon appears at $\omega$ = 600 cm$^{-1}$, which is beyond the accessible range of our lasers. Modeling also shows that the feature in the phase spectra $\phi(\omega)$ originating from the Dirac plasmon occurs at higher frequency compared to the amplitude spectra, and can therefore be probed by our experimental setup (Fig. 2d). We attribute the observed increase of the phase at low frequencies (Fig. 2c) to direct near-field coupling to the Dirac plasmon. This finding, along with the fingerprints of plasmon-phonon interaction, establishes graphene as a new medium supporting plasmonic effects. Unlike noble metals: traditional materials supporting surface plasmons, graphene is inherently tunable by electric and magnetic fields, thus enabling functionalities not attainable with metal plasmonics.

The combination of high-momentum spectroscopy and nano-imaging demonstrated in our work sets the stage for studying many other properties of Dirac plasmons in graphene. Of special interest are effects pertaining to the real-space confinement and propagation of plasmons in nano-structures/ribbons[35,36]. A modification of the plasmon dispersion and/or ultra-fast modulation[37] of the Dirac plasmon can be conveniently carried out through back-gating with a degree of control that is difficult to obtain within all-metal plasmonics. Turning to the high-$q$ spectroscopy aspects of tip-induced light-matter interaction, we wish to point out that a much broader range of $q$ may be interrogated using super sharp silicon tips ($a$ < 10 nm) and even sharper tips based on carbon nano-tubes ($a$ down to 1 nm)[38]. Such a further expansion of the momentum space accessible by IR nanoscopy, combined with the improved spatial resolution, is especially appealing in the context of studying collective modes in the vicinity of the single-particle excitation continuum, and manipulating light in graphene-based nanostructures or transformation optics elements.

## ■ ASSOCIATE CONTENTS

Supporting Information Available: Supporting experimental data and theory details. This material is available free of charge via the Internet at http://pubs.acs.org.

## ■ AUTHOR INFORMATION

Corresponding Author

*Email: zfei@physics.ucsd.edu and gandreev@physics.ucsd.edu.

Author Contributions

*These authors contributed equally to this work.

## ■ ACKNOWLEGEMENT

This work is supported by ONR, DOE, AFOSR, NASA, MURI and Deutsche Forschungsgemeinschaft through the Cluster of Excellence Munich Centre for Advanced Photonics.

# ■ NOTES ADDED IN PROOF

After submitting this manuscript we became aware of the work by L. Ju et al., where the authors report on plasmonic effects in the graphene micro-ribbons in the the terahertz frequencies[39].

# ■ REFERENCE